\documentclass[fleqn,10pt]{wlscirep}
\title{A Single-Ion Reservoir as a High-Sensitive Sensor of Electric Signals }
\author[1]{Francisco Dom\'inguez}
\author[2]{I\~nigo Arrazola}
\author[1]{Jaime Dom\'enech}
\author[2]{Julen S. Pedernales}
\author[2]{Lucas Lamata}
\author[2,3]{Enrique Solano}
\author[1,4,*]{Daniel Rodr\'iguez}

\affil[1]{Departamento de F\'isica At\'omica, Molecular y Nuclear, Universidad de Granada, Campus de Fuentenueva s/n, 18071, Granada, Spain}
\affil[2]{Department of Physical Chemistry, University of the Basque Country UPV/EHU, Apartado  644, 48080 Bilbao, Spain}
\affil[3]{IKERBASQUE, Basque Foundation for Science, Maria Diaz de Haro 3, 48013 Bilbao, Spain}
\affil[4]{Centro de Investigaci\'on en Tecnolog\'ias de la Informaci\'on y las Comunicaciones,  Universidad de Granada, 18071, Granada, Spain}
\affil[*]{danielrodriguez@ugr.es}

\begin{abstract}
A single-ion reservoir has been tested, and characterized in order to be used as a highly sensitive optical detector of electric signals arriving at the trapping electrodes. Our system consists of a single laser-cooled $^{40}$Ca$^+$ ion stored in a Paul trap with rotational symmetry. The performance is observed through the axial motion of the ion, which is equivalent to an underdamped and forced oscillator. Thus, the results can be projected also to Penning traps. We have found that, for an ion oscillator temperature  $T_{\scriptsize{\rm axial}}\lesssim 10$~mK in the forced-frequency range $\omega _z =2\pi \times (80,200$~kHz), the reservoir is sensitive to a time-varying electric field equivalent to an electric force of $5.3(2)$~neV/$\mu $m, for a measured quality factor $Q=3875(45)$, and a decay time constant $\gamma _z=88(2)$~s$^{-1}$.  This method can be applied to measure optically the strength of an oscillating field or induced (driven) charge in this frequency range within times of tens of milliseconds. Furthermore the ion reservoir has been proven to be sensitive to electrostatic forces by measuring the ion displacement. Since the heating rate is below $0.3$~$\mu$eV/s, this reservoir might be used as optical detector for any ion or bunch of charged particles stored in an adjacent trap.
\end{abstract}
\begin{document}

\flushbottom
\maketitle
%
%
\thispagestyle{empty}

\section*{Introduction}

Single-ion sensitivity in a trap is well-reached by applying laser cooling and by optically observing electronic transitions in atomic ions \cite{Neuh1980,Itan1981}. This is prominently used in experiments and proposals devoted to simulate quantum systems~\cite{Blat2012, Arrazola16}, to process quantum information~\cite{Haffner08,Pedernales14}, or to implement optical clocks \cite{Schm2005}. Experiments on a single laser-cooled ion, or on a set of them addressing each one individually, are routinely carried out in Paul \cite{Leib2003} and Penning traps~\cite{Mava2014}. Optical detection has permitted even to observe subtle effects like the interaction between two ions stored in different potential wells of the same trap structure, through their motional phonons \cite{Brow2011}. 

In this article, we will demonstrate the advantage of using a laser-cooled ion for the measurement of low-amplitude electric signals, with a  potential application in Penning-trap mass spectrometry~\cite{Blau2013}.  The Penning-trap techniques utilized to reach single-ion sensitivity are based on the detection of the current induced by the stored ion on the trap electrodes, which has to be resonantly amplified and electronically filtered to yield the cyclotron frequency \cite{Wine1975}. Despite the success from the implementation of different variants of the technique \cite{Corn1989,Haef2003,Vand2006}, e.g. the recent ultra-accurate measurements of the masses of the electron, and the proton-to-antiproton mass-to-charge ratio \cite{Stur2014,Ulme2015}, the highest sensitivity is still limited to charged particles with low or medium mass-to-charge ratios. These are practicable if they oscillate with large energies in the potential well of the trap. Although the dynamics of a laser-cooled ion is well known \cite{Leib2003}, a single ion has never been used as an ion-motion detector. Substituting the electronic circuit placed in a liquid-helium tank by a laser-cooled atomic ion will yield an improvement in sensitivity, with the prospect also to reach better accuracy. The need of a mass-independent single-ion detection method is of interest to perform direct measurements of the binding energies of superheavy elements \cite{Bloc2010}. Moreover, a better accuracy in the determination of the atomic mass would be useful for the determination of the mass of the electron neutrino at the sub-eV level \cite{Elis2015}.

The use of a Doppler-cooled $^{40}$Ca$^+$ as a high-sensitive and fast photon detector for precise mass measurements was presented in Ref.~\cite{Rodr2012}, relying on the coupling between two ions, a concept proposed earlier by~D.~J.~Heinzen and D. J.~Wineland \cite{Wine1990}. In that scenario, the energy transferred by the ion under investigation to its partner is quantified through the fluorescence photons emitted by the latter. The performance of the laser-cooled $^{40}$Ca$^+$ ion has been studied both experimentally and theoretically. We will consider the distributions of the emitted photons by this ion, within a model that is based on a damped-forced harmonic oscillator. Here, the force comes from a time-varying electric dipole field directly applied by the trapping electrodes. We will also discuss the effect of applying an electrostatic potential and white noise.

\section*{Results}

The experiments have been carried out using the open-ring Paul trap described in Ref.~\cite{Corn2015}, but now driving the trapping field with a radiofrequency $\omega _{\scriptsize{\hbox{RF}}}=2\pi \times 1.47$~MHz and applying a compensation DC voltage ($U_{\scriptsize{\hbox{DC}}}$) to some of the trap electrodes. The $^{40}$Ca$^+$ ion is created by photoionization at a background pressure of $\approx 10^{-10}$~mbar. The level structure of  $^{40}$Ca$^+$  shown in Fig.~\ref{fig:new_1} contains two electric dipole-allowed transitions, which will contribute to the dynamics.  Three laser beams have been used for this experiment as shown schematically in the right-hand side of Fig.~\ref{fig:new_1}, with tunable wavelengths around 397~nm (B1 and B2), and 866~nm (R1). B1 and B2 are tuned to drive the cooling transition 4s$^2$S$_{1/2}\rightarrow$4s$^2$P$_{1/2}$ in the radial and axial directions, respectively. Moreover, R1, which is only applied in the radial direction, is needed for the transition 3d$^2$D$_{3/2}\rightarrow$4s$^2$P$_{1/2}$ to pump the metastable state 3d$^2$D$_{3/2}$, unavoidably populated with a probability of 7$\%$. The natural linewidths of  the 3d$^2$D$_{3/2}\rightarrow$4s$^2$P$_{1/2}$ and 4s$^2$S$_{1/2}\rightarrow$4s$^2$P$_{1/2}$ transitions are $\Gamma _{\scriptsize{\hbox{R}}}= 2\pi \times 1.35$~MHz, and  $\Gamma _{\scriptsize{\hbox{B}}}= 2\pi \times 21.58$~MHz, respectively. The fluorescence light from the cooling transition (397~nm) is collimated using a commercial system from Thorlabs (MVL12X12Z), which provides a magnification of $\simeq 6.75(5)$ in the focal plane of an EMCCD (Electron Multiplier Charged Couple Device) from Andor (IXON3). The EMCCD sensor is made of 512 $\times$ 512 pixels. The size of each pixel is $16$~$\mu $m $\times$ $16$~$\mu $m. An interference filter is located in front of the EMCCD to collect only photons with wavelength around 397~nm. The effective pixel size has been obtained through the ratio between the minimum distance, due to Coulomb interaction between two trapped ions $\Delta z=(e^2/2\pi \epsilon _0m\omega _{{\scriptsize{\hbox{z}}}}^2)^{1/3}$ \cite{Jame1998}, and the pixel distance between the centers of the distributions of the ions projected on the axial direction. The radiofrequency voltage ($V_{\scriptsize{\hbox{RF}}}$) has been varied from $\approx 520$ to $1430$~V$_{\scriptsize{\hbox{pp}}}$.  The trap frequencies are given by 
\begin{equation}
\omega _{u}=\frac{\omega _{\scriptsize{\hbox{RF}}}}{2}\sqrt{a _{u}+\frac{q  _{u}^{\scriptsize{\hbox{2}}}}{2}},   \qquad u=r,z
\end{equation}
where $q  _{u}=f(V_{\scriptsize{\hbox{RF}}},\omega _{\scriptsize{\hbox{RF}}},u _{\scriptsize{0}})$ and $a_{\scriptsize{\hbox{u}}}=f(U_{\scriptsize{\hbox{DC}}},\omega _{\scriptsize{\hbox{RF}}}, u _{\scriptsize{0}})$ with $u_{\scriptsize{0}}$ representing the characteristic dimensions of the trap \cite{Ghos1995}. The trap frequencies in the axial direction $\omega _{\scriptsize{\hbox{z}}}$ were measured to be in the range $\approx 2\pi \times 60$-$250$~kHz, and the corresponding trap frequencies in the radial direction  $\omega _{\scriptsize{\hbox{r}}}$ are a factor of two smaller provided no DC potential is applied. $q_{\scriptsize{\hbox{z}}}$ in the axial direction varies from 0.18 to 0.49 for different values of $V_{\scriptsize{\hbox{RF}}}$ and  $a_{\scriptsize{\hbox{z}}}=-0.0088$ for the measurements carried out with $\omega _{\scriptsize{\hbox{z}}}\simeq 2\pi \times 108$~kHz($q _{\scriptsize{\hbox{z}}} \simeq 0.25$). This frequency is close to the value originally proposed~\cite{Rodr2012} and within a range where the adiabatic approximation is valid~\cite{Ghos1995}. For the sensitivity measurements to electrostatic potentials,  $a_{\scriptsize{\hbox{z}}}=-0.0184$ and $\omega _{\scriptsize{\hbox{z}}}\simeq 2\pi \times 80$~kHz for the same $q _{\scriptsize{\hbox{z}}}$.

Our trapped ion can be considered, to a good approximation, a harmonic oscillator in all three directions.  The effect of the described laser configuration will be modeled as a damping force acting on the direction of the cooling lasers, which is the standard treatment in the theory of Doppler cooling~{ \cite{Leib2003,Stenholm1986,Akerman2010}. As a consequence, the dynamics of the ion in the axial direction is given by a damped harmonic oscillator, with equation of motion
\begin{equation}
\ddot \rho_z +2 \gamma_z \dot \rho_z + \omega^2_z \rho_z =0, \label{osci}
\end{equation}
where $\rho_z$ and $\gamma_z$  are, respectively,  the position of the ion and the damping coefficient in the axial direction. A term given by $\alpha \rho_z^3$, might be added to the left part of Eq.~(\ref{osci}) to account for possible deviations of the quadrupole potential due to the DC voltages applied for compensation.

When the ion reaches a temperature given by the Doppler limit, the cooling effect  of the lasers stops and the ion reaches a steady state. In a three-level system, this limit temperature will depend on the intensities and frequencies of the lasers driving the cooling and pumping transitions. In this steady state, the ion is sensitive to external oscillating electric fields acting on the electrodes of the trap.  This signal could be originated by one or more ions in a second trap. However, in all experiments carried out here, this electric field is artificially generated. This will test the sensitivity of our oscillator, which ultimately would act as a weighing device to measure the mass of one or many ions in the other trap~\cite{Wine1990}. 

In Fig.~\ref{fig:new_2}a,  we show the distribution of photons emitted by the ion as it oscillates in the axial direction while in its steady state. We find that the width of the photon distribution is related to the variance of the projected Gaussian-like distribution, and in this way, to the amplitude of the oscillation $\rho_{z, \rm max}$. To explain this, we assume that the photons emitted by the ion at each point in the axial line are collected at the CCD camera with a Lorentzian distribution. The width of this distribution is obtained from a $\chi ^2$ fit to the profiles of the fluorescence measurements for different oscillation amplitudes when applying a driving force as shown in Fig.~\ref{fig:new_2}b for $\rho_{z, \rm max}=21~{\rm \mu m}$. The function for the $\chi^2$ fit is built by the convolution of the probability distribution of the oscillator in the axial direction $P(\rho_z)$, and the Lorentzian distribution, considering on one hand $\rho_{z, \rm max}$, and on the other hand the Lorentzian width, as free parameters.  This demonstrates that we can optically detect the effect of a time-varying electric signal applied to the electrodes of the trap while we obtain the optical response of our system. The uncertainties in $\rho_{z, \rm max}$ are the quadratic sum of the statistical uncertainty (from one measurement), and the systematic uncertainties, arising from the election of the position of the central pixel,  from the resolution and magnification of the imaging system, and from the step size in $\rho_{z, \rm max}$. We will now systematically use this detection method to calibrate the sensitivity of our device. Additionally, we can associate an axial temperature to the ion in equilibrium according to  $k_{\scriptsize{\hbox{B}}}T_z \equiv m \langle v^2_z \rangle =m \omega_z^2  \langle \rho^2_z \rangle= m \omega_z^2 \sigma^2_z$~\cite{Leib2003}. In this manner, we measure an interval of Doppler limit temperatures for the trap frequencies tested in this work from 8.1(3.8) to 10.7(2.9)~mK. The same can be done for the radial direction, provided one knows the radial frequency. 
In order to characterize the sensitivity of our single-ion reservoir, we measure its response to dipolar fields of different amplitudes and frequencies, as shown in Fig.~\ref{fig:new4}.  In the steady state, the amplitude of the driven-damped harmonic oscillator is given by 
\begin{equation}
\rho _{z,\scriptsize{\hbox{max}}}=\frac{F_e}{m} \{ (2 \gamma_z \omega_{\rm dip})^2 + (\omega_z^2 - \omega_{\rm dip}^2)^2 \}^{-1/2}, \label{fit}
\end{equation}
where $F_e$ and $\omega_{\rm dip}$ are , respectively, the amplitude and the frequency of the harmonic driving force.
There is no analytical expression for $\rho _{z,\scriptsize{\hbox{max}}}$ if one introduces the anharmonicity term $\alpha \rho^3_z$. In such case however, it is possible to perform a numerical fit. Fitting the experimental data ($V_2$ in Fig.~\ref{fig:new4}) with Eq.~(\ref{fit}), we obtain $F_{e,2}=5.3$(2) neV/$\mu$m, and a damping coefficient $\gamma_{z,2} = 88(2)$~Hz, to which a quality factor $Q=\frac{\omega _z}{2 \gamma_z}=3875(45)$ can be associated. Introducing a term to account for an anharmonicity and performing a numerical fit, one obtains values consistent with the latter. In case the amplitude of the external field is $V_3$, we obtained $F_{e,3}=3.6(1)$~ neV/$\mu$m. The ratio of the electric field the ion sees $F_{e,3}/F_{e,2}=0.68(5)$ equals the ratio of the amplitudes applied to the electrodes of $V_3/V_2=0.7$, thus by reducing this amplitude, it is possible to infer an electric field on the ion of $0.1F_{e,2}$, when a radiofrequency field with an amplitude $V_4=0.1V_2$ is applied. The resonance curve is shown for a frequency of $2\pi \times 202$~kHz in the lower-right size of Fig.~\ref{fig:new4}. The signal-to-noise ratio is $\sim 2$. 

The application of this ion reservoir to measure ion-motional frequencies is subject to the possibility of observing fluorescence signals in short periods of time. Figure~\ref{fig:10}a shows the image collected for different delay times after the ion has been excited. Each image is collected during 20~ms, which gives a good signal-to-noise ratio without affecting the time resolution. The observed decay rates applying different amplitudes of the driving field are in agreement with the ones computed from the fitting of Fig.~\ref{fig:new4} ($\gamma _{z,2}$). 

We have demonstrated that the resonance of the ion is clearly visible even in cases when the external field amplitude is the lower available in our laboratory ($V_4$  in Fig.~\ref{fig:new4}). This field should be equivalent to a force of 0.53 neV/$\mu $m, in contrast to the force that a single ion stored in an adjacent trap would originate, which is of $F_e=0.01$~neV/$\mu$m, considering an electronic charge state of 1$^+$ and an oscillation amplitude $\rho _{z,\scriptsize{\hbox{max}}}=100$~$\mu $m in the micro-trap described in Ref.~\cite{Corn2016}. The minimum temperature reached in these measurements corresponds to a mean number of $\sim1900$~phonons in the axial direction ($E _{\scriptsize{\hbox{phonon}}}=4.5$~neV for $\omega _z=2 \pi \times 108$~kHz).  Using the scheme based on fluorescence photons depicted in Fig.~\ref{fig:10}, one needs to couple the motion of the $^{40}$Ca$^+$ ion with the motion of the ion to be studied and observe the evolution of the system in the absence of lasers, i.e. without cooling \cite{Rodr2012}. The left part of Fig.~\ref{fig:10}b shows the image in the sensor and the axial projection when an excitation has been previously applied, in the absence of lasers, during 50~ms. From the Gaussian fit to the axial distribution, and considering $\gamma_z = 88(2)$~Hz, one obtains, after excitation, an initial ion oscillation amplitude  $\rho_{\scriptsize{\hbox{z,max}}} = 15.6$~$\mu$m ($T_z = 230$~mK). This value, compared with the result when no excitation is applied, while there is no interaction with the laser beams, yields a rate for the increase in energy of the order of 10~$\mu$eV/s, much larger than the measured heating rate, below $0.3$~$\mu$eV/s. This upper limit for the heating rate has been obtained by measuring the response of the laser-cooled trapped ion  to different amplitudes of applied white noise ($V_{\scriptsize{\hbox{noise}}}$) and fitting linearly the axial temperature with the function given by

\begin{equation}
T_z=\frac{1}{\gamma _z k_B}(K+\zeta\cdot V_{\scriptsize{\hbox{noise}}}^2),
\end{equation}

where $\zeta\cdot V^2$ is the heating rate in units of J s$^{-1}$, which is multiplied by the square of the amplitude of the noise signal and $K$ is a constant from which we recover the Doppler limit, in the ideal case where $V_{\rm noise}=0$. Then, the square of the variance $\sigma _z^2(t)=2\zeta V_{\rm noise}^2/(m\omega _z ^2)\cdot t$.  The upper limit obtained is just the value corresponding to the smallest amplitude applied.

For completeness, we have also studied the response of the system to DC fields. The electrostatic force in the axial direction is given by $F_z=m\omega _z^2 z$. Figure~\ref{fig:6} shows the electrostatic force as a function of the ion displacement, which is defined as the center of the Gaussian distributions resulting from the projections of the fluorescence images in the axial direction (see the insets). From the fit, one can get $F_z/z=0.10455(0)$~neV/$\mu$m/nm. The center of the distributions is determined with an uncertainty of $67$~nm, which defines the uncertainty in the measurement of the electrostatic force of 7~neV/$\mu$m. The minimum force measured from the ion displacement in the axial direction shown in Fig.~\ref{fig:6}, marked with (a) is 42(7)~neV/$\mu$m. Smaller displacements have been obtained in the radial direction but the radial frequency was not measured.

\section*{Discussion}

Summarizing, this article experimentally realizes an atomic ion reservoir composed of a trapped ion as a high-sensitive measurement device. The Doppler-cooled ion will allow one to detect weak electric fields as well as the motional frequencies of charged particles in adjacent traps.  Our experiment paves the way towards more sensitive mass and frequency spectroscopy with charged particles. This method can be extended to a broader frequency range, where ground-state cooling on the same ion can be applied. This might allow increasing the sensitivity of the optical method in the DC and radiofrequency regimes, as well as controlling and measuring the phase of the axial motion.

\section*{Acknowledgements}

We acknowledge support from the European Research Council (ERC StG contract 278648-TRAPSENSOR); Spanish MINECO/FEDER FPA2012-32076, FPA2015-67694-P, UNGR10-1E-501 and FIS2015-69983-P; Ram\'on y Cajal Grant RYC-2012-11391; Basque Government PhD grant PRE-2015-1-0394 and project IT986-16; UPV/EHU PhD grant and UPV/EHU UFI 11/55. F.D. acknowledges support from Spanish MINECO ``Programa de Garant\'ia Juvenil'' cofunded by the University of Granada. We warmly thank Ra\'ul Rica for fruitful discussions on thermal motion of laser-cooled trapped ions.

\section*{Author contributions statement}

F.D., J.D. and D.R. performed the experiments. I.A., J.S.P., L.L. and E.S. provided theoretical physics support. F.D., I.A., J.S.P., L.L., E.S. and D.R. contributed to the generation and development of the ideas and to the writing of the paper.

\section*{Additional information}

\textbf{Competing financial interests:} The authors declare no competing financial interests.

\begin{figure}[ht]
\centering
\includegraphics[width=0.7\linewidth]{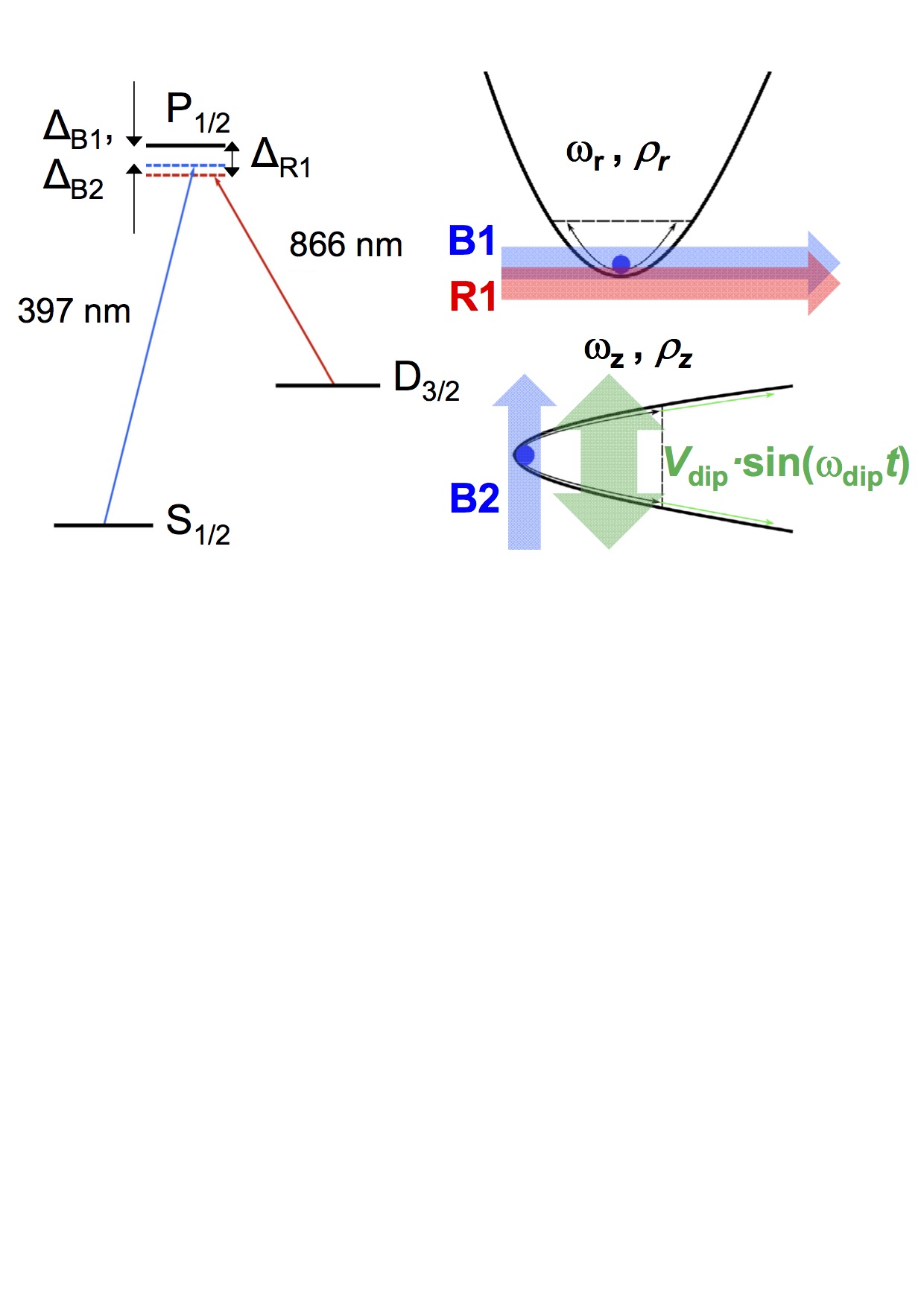}
\caption{{\bf Atomic levels and lasers configuration to perform Doppler cooling on $^{40}$Ca$^+$:} 
$^{40}$Ca$^+$ atomic levels for Doppler cooling and schematic representation of the interaction between a trapped ion and laser beams in the radial and axial potential wells of the open-ring trap \cite{Corn2015}. The double-ended thick arrow depicts the application of an external dipole field with amplitude $V_{\scriptsize{\hbox{dip}}}$ equivalent to a force $F_e/m$, where $m$ is the mass of the ion. \label{fig:new_1}}
\end{figure}

\begin{figure}[ht]
\centering
\includegraphics[width=0.7\linewidth]{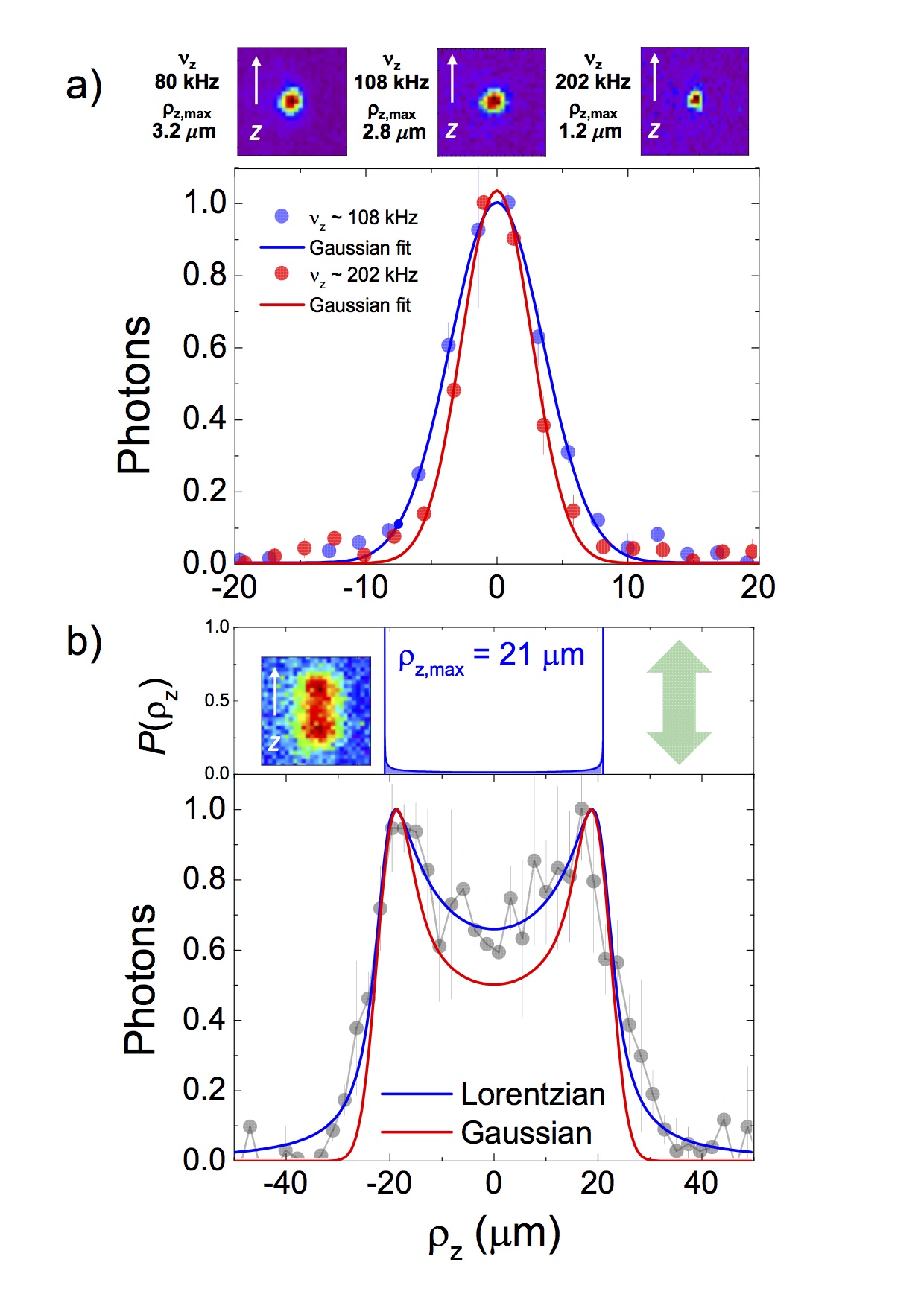}
\caption{{\bf The analysis and evaluation method:}
(a) Fluorescence images for different trap frequencies and normalized photon distributions in the axial direction for $\nu _z=108,202$~kHz. b)  Probability density $P(\rho _z)$ and normalized photon distributions when applying continuously an external time-varying dipole field in the axial direction. In the presence of the laser field, the ion oscillates with a constant amplitude. The signals are built averaging the detected photons in 2-3 rows of pixels in the radial direction.} \label{fig:new_2}
\end{figure}

\begin{figure}[ht]
\centering

\includegraphics[width=0.75\linewidth]{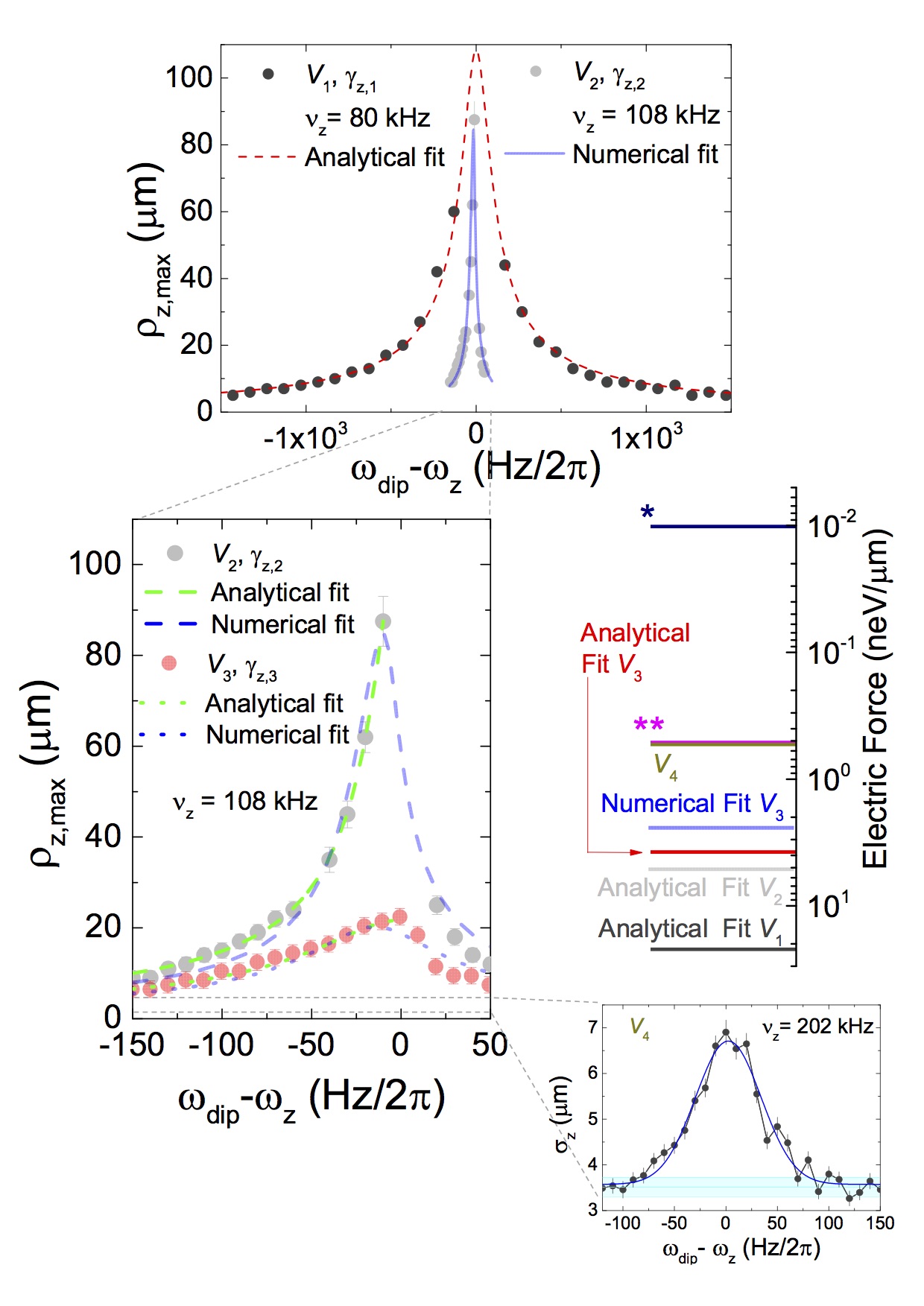}
\caption{{\bf Radiofrequency force sensing:} 
Upper: $\rho _{z,\scriptsize{\hbox{max}}}$ as a function of $(\omega_{\scriptsize{\hbox{dip}}}-\omega _z)/2\pi $ for two amplitudes $V_{\scriptsize{\hbox{dip}}}$ of the time-varying dipole field and different oscillation frequencies. The laser beams are always interacting with the ion. The analytical fit is carried out using Eq.~(\ref{fit}), while the numerical fit takes into account the anharmonicity. Middle left: zoomed plot showing the data taken for $\omega _z=2\pi \times 108$~kHz for different amplitudes,  i.e., $V_2\approx 125$~$\mu $V$_{\scriptsize{\hbox{pp}}}$  and $V_3 = 0.7$$V_2$, and different laser parameters, resulting in $\gamma_{z,2} = 88(2)$~Hz and $\gamma_{z,3} = 298(24)$~Hz, respectively, from the analytical fit. Middle right: sensitivity of the ion reservoir to electric forces in neV/$\mu $m. $^*$ shows the field generated by a single-charged ion oscillating in a second trap, and $^{**}$ the field generated by e.g. 50 antiprotons or 50 single-charged superheavy-element ions. The lower right side of the figure, shows the variance of the ion distribution as a function of $(\omega_{\scriptsize{\hbox{dip}}}-\omega _z)/2\pi $ for an amplitude $V_4=0.1V_2$. The cyan-colored area shows the noise level, which is defined by the average variances of the distributions off resonance. Since the invariance for each data point is obtained directly from the Gaussian fit to the projection of the fluorescence into the axial direction, the uncertainty of each data point is the quadratic sum of the statistical and systematic uncertainties, the first provided by the fit and the second by the observed variations. \label{fig:new4}}
\end{figure}

\begin{figure}[ht]
\centering
\includegraphics[width=0.75\linewidth]{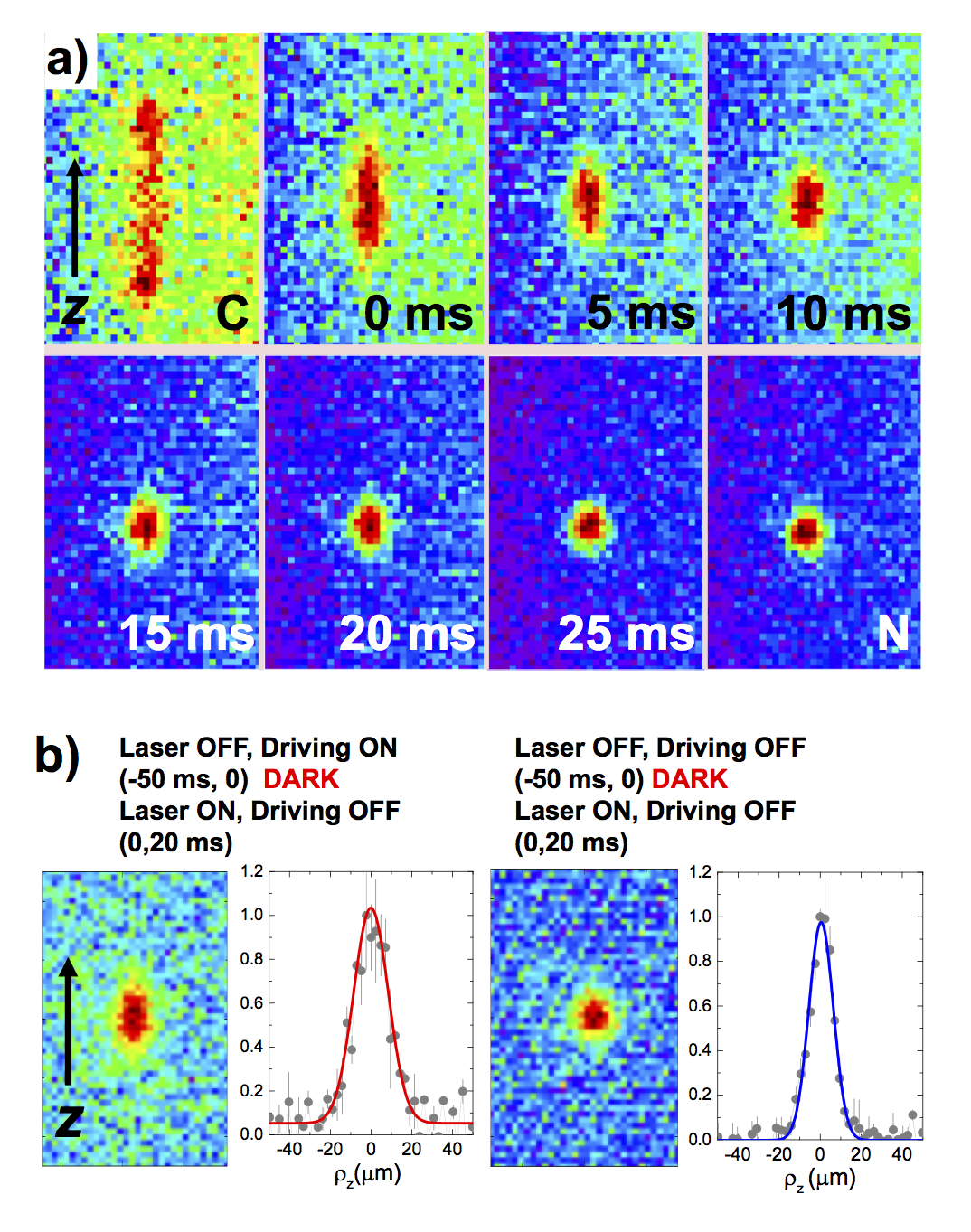}
\caption{{\bf EMCCD pictures taken in trigger mode:} 
(a) Fluorescence images, collected during 20~ms when an external field is applied continuously (C), when the start time to trigger the acquisition is delayed from 0 to 25~ms with respect to a time zero (external field is removed), and when no external field is applied (N). Each picture is the sum of 200~cycles. (b) Images collected during 20~ms and projections in the axial line when the acquisition is triggered on switching ON the laser R1.\label{fig:10}}
\end{figure}

\begin{figure}[ht]
\centering
\includegraphics[width=0.7\linewidth]{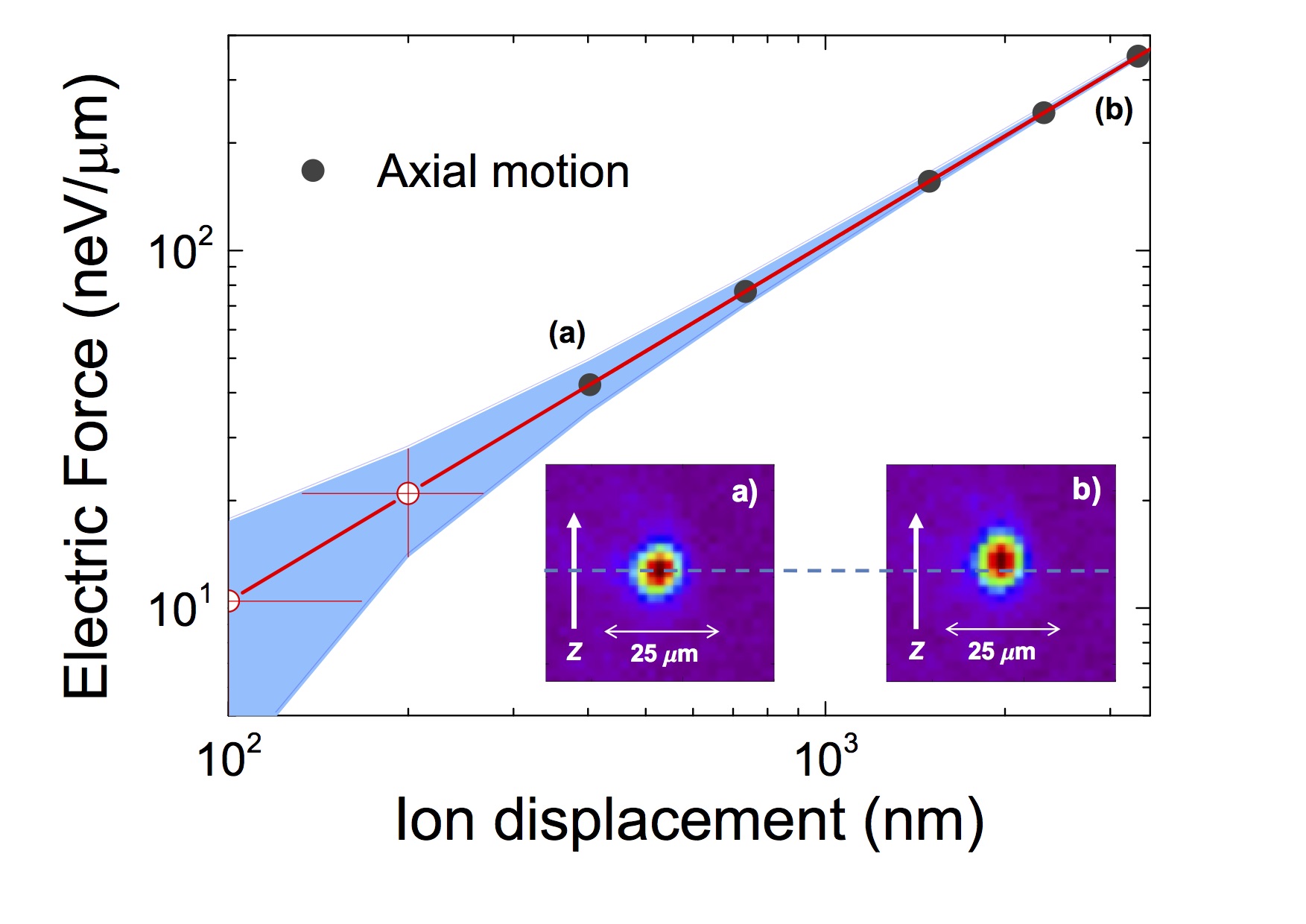}
\caption{{\bf Electrostatic force sensing:}  Electrostatic force exerted on the laser-cooled ion as a function of the ion displacement. The potential applied to a specific electrode was varied from 0.5~V in (a) to 4~V in (b). The axial frequency was 80~kHz. The inset shows the images collected for the cases a) and b). Each of the images was taken during 35~seconds. The red line is a linear fit to the data points (black-solid circles), and the blue area delimit the uncertainties. The white-solid circles are obtained after the fit. \label{fig:6}}
\end{figure}


\begin{thebibliography}{9}

\bibitem{Neuh1980} Neuhauser, W., Hohenstatt, M., Toschek, P. E., \& Dehmelt, H. G. Localized visible Ba$^+$ mono-ion oscillator. \emph{Phys. Rev. A} {\bf 22}, 1137 (1980). 

\bibitem{Itan1981} Wineland, D. J., \& Itano,W. M. Spectroscopy of a single Mg$^+$ ion. \emph{Phys. Lett. A} {\bf 82}, 75 (1981).

\bibitem{Blat2012}  Blatt, R. \& Roos, C.  Quantum simulations with trapped ions. \emph{Nature Phys.} {\bf 8}, 277 (2012).

\bibitem{Arrazola16} Arrazola, I., Pedernales, J. S., Lamata, L., \& Solano, E. Digital-Analog Quantum Simulation of Spin Models in Trapped Ions. \emph{Sci. Rep.} {\bf 6}, 30534 (2016).

\bibitem{Haffner08} H\"affner, H., Roos, C. F.,  \& Blatt, R.  Quantum computing with trapped ions. \emph{Phys. Rep.} {\bf 469}, 155 (2008).

\bibitem{Pedernales14} Pedernales, J. S., \emph{et al.} Entanglement Measures in Ion-Trap Quantum Simulators without Full Tomography. \emph{Phys. Rev. A} {\bf 90}, 012327 (2014).

\bibitem{Schm2005} Schmidt, P. O., \emph{et al.} Spectroscopy Using Quantum Logic. \emph{Science} {\bf 309}, 749 (2005).

\bibitem{Leib2003} Leibfried, D., Blatt, R., Monroe, C., \& Wineland, D. Quantum dynamics of single trapped ions. \emph{Rev. Mod. Phys.} {\bf 75}, 281 (2003).

\bibitem{Mava2014} Mavadia, S., \emph{et al.} Optical sideband spectroscopy of a single ion in a Penning trap. \emph{Phys. Rev. A} {\bf 89}, 032502 (2014).

\bibitem{Brow2011} Brown, K. R., \emph{et al.} Coupled quantized mechanical oscillators. \emph{Nature} {\bf 471}, 196 (2011).

\bibitem{Blau2013} Blaum, K., Dilling, J., \& N\"ortersh\"auser, W. Precision atomic physics techniques for nuclear physics with radioactive beams. \emph{Phys. Scr.} {\bf T152}, 014017 (2013).

\bibitem{Wine1975} Wineland, D. J. \& Dehmelt, H. G. Principles of the stored ion calorimeter. \emph{J. Appl. Phys.} {\bf 46}, 919 (1975).

 \bibitem{Corn1989} Cornell, E. A., \emph{et al.} Single-ion cyclotron resonance measurement of M(CO$^+$)/M(N$^+_2$). \emph{Phys. Rev. Lett.} {\bf 63}, 1674 (1989).

\bibitem{Haef2003} H\"affner, H., \emph{et al.} Double Penning trap technique for precise g factor determinations in highly charged ions. \emph{Eur. Phys. J. D}{\bf 22}, 163 (2003).

\bibitem{Vand2006} Van Dyck, R. S. Jr., Pinegar, D. B., Liew, S. V., \& Zafonte, S. L. The UW-PTMS: Systematic studies, measurement progress, and future improvements.  \emph{Int. J.  Mass Spectrom.} {\bf 251}, 231 (2006).

\bibitem{Stur2014} Sturm S., \emph{et al.} High-precision measurement of the atomic mass of the electron. \emph{Nature} {\bf 506}, 467 (2014).

\bibitem{Ulme2015} Ulmer, S., \emph{et al.} High-precision comparison of the antiproton-to-proton charge-to-mass ratio. \emph{Nature} {\bf 524}, 196 (2015).

\bibitem{Bloc2010} Block, M., \emph{et al.} Direct mass measurements above uranium bridge the gap to the island of stability. \emph{Nature} {\bf 463}, 785 (2010).

\bibitem{Elis2015} Eliseev, S., \emph{et al.} Direct Measurement of the Mass Difference of $^{163}$Ho and $^{163}$Dy Solves the Q-Value Puzzle for the Neutrino Mass Determination. \emph{Phys. Rev. Lett.} {\bf 115}, 062501 (2015).

\bibitem{Rodr2012} Rodr\'iguez, D. A quantum sensor for high-performance mass spectrometry. \emph{Appl. Phys. B} {\bf 107}, 1031 (2012).

\bibitem{Wine1990} Heinzen, D. J. \& Wineland, D. J., Quantum-limited cooling and detection of radio-frequency oscillations by laser-cooled ions. \emph{Phys. Rev. A} {\bf 42}, 2977 (1990).

\bibitem{Corn2015} Cornejo, J. M., \emph{et al.} Extending the applicability of an open-ring trap to perform experiments with a single laser-cooled ion.  \emph{Rev. Sci. Instrum.} {\bf 86}, 103104 (2015).

\bibitem{Jame1998} James, D. F. V. Quantum dynamics of cold trapped ions with application to quantum computation. \emph{Appl. Phys. B} \textbf{66}, 181 (1998).

\bibitem{Ghos1995} Ghosh, P. K. Ion Traps. \emph{Clarendon Press} (Oxford), (1995).

\bibitem{Stenholm1986} Stenholm, S. The semiclassical theory of laser cooling. \emph{Rev. Mod. Phys.} {\bf 58}, 699 (1986).

\bibitem{Akerman2010} Akerman, N., \emph{et al.} Single-ion nonlinear mechanical oscillator. \emph{Phys. Rev. A} {\bf 82}, 061402(R) (2010).

\bibitem{Corn2016} Cornejo, J. M., \emph{et al.} An optimized geometry for a micro Penning-trap mass spectrometer based on interconnected ions. \emph{Int. J. Mass Spectrom.} {\bf 410C}, 22 (2016).

\end{thebibliography}
\end{document}